\documentclass[12pt]{article}
\usepackage{cite}
\usepackage{graphicx}
\usepackage[utf8]{inputenc}
\usepackage[dvips]{epsfig}
\usepackage{graphicx}
\usepackage[T1]{fontenc}
\usepackage{xcolor}
\usepackage{color,soul}
\usepackage{authblk}
\usepackage{amsmath}
\usepackage{float}
\usepackage{algorithm}
\usepackage{algpseudocode}
\usepackage{hyperref}
\usepackage{bbold}
\date{}
\usepackage{algorithm}

\usepackage{subcaption}

\date{\today}

\title{\bf Scalable Entanglement Detection in Quantum Systems via Fisher Linear
	Discriminant Analysis }
\author[1,2,3]{Mahmoud Mahdian\thanks{mahdian@tabrizu.ac.ir}}
\author[1]{Zahra Mousavi\thanks {z.mosavi1400@ms.tabrizu.ac.ir}}
\date{\today}

\affil[1]{Faculty of Physics, Theoretical and Astrophysics
	Department, University of Tabriz, 51665-163 Tabriz, Iran}
\affil[2]{Research Institute for Applied Physics and Astronomy (RIAPA), University of Tabriz, Tabriz, Iran}
\affil[3]{Quantum Technology Center, University of Tabriz, Tabriz, Iran}

\begin{document}
	\maketitle
	\begin{abstract}
		
		Quantum entanglement is the cornerstone of quantum technology and enables quantum devices to outperform classical systems in terms of performance. However, detecting entanglement in high-dimensional systems remains a significant challenge due to the exponential growth of the Hilbert space with the number of particles. In this work, we use machine learning to classify entangled states and separable states, focusing on the application of classical Fisher Linear Discriminant Analysis (FLDA). By adapting classical statistical learning techniques to quantum state discriminant analysis, we present the theoretical foundations, a practical implementation strategy, and the advantages of FLDA in this context. We systematically evaluate the performance of this method on different quantum states and demonstrate its effectiveness as a tool for efficient quantum state classification. Finally, we investigate multi-qubit quantum states with high accuracy and classify these states.
	\end{abstract}
	
	\section{Introduction}

Quantum entanglement is a defining characteristic of quantum mechanics, enabling correlations between particles that have no classical counterpart \cite{einstein1935can, bell1964einstein, horodecki2009quantum}. As a critical resource in quantum computing \cite{nielsen2010quantum, preskill2018quantum, knill2001quantum}, communication \cite{bennett1993teleporting, pan2012multiparticle, pirandola2020advances}, and sensing \cite{degen2017quantum, guhne2009entanglement, giovannetti2011advances}, entanglement is fundamental to achieving a quantum advantage. However, detecting and characterizing entanglement especially in high-dimensional or multipartite systems remains a significant challenge \cite{guhne2009entanglement, friis2019entanglement, pan2012multiparticle, briegel2021entanglement, lu2020quantum, shang2021efficient, li2022highdimensional, zhang2023scalable, huang2024adaptive, chen2023efficient, wang2024multipartite}. Traditional methods, such as entanglement witnesses \cite{horodecki2009quantum, guhne2009entanglement, kim2023advances}, quantum state tomography \cite{nielsen2010quantum, friis2019entanglement, yang2022tomography}, or Bell inequality violations \cite{bell1964einstein, augusiak2014bell, brunner2014bell}, often require precise knowledge of the state, extensive measurements, or exponential computational resources. This makes scalable entanglement detection an open problem in quantum information science \cite{yu2022scalable, zhang2023scalable, wang2024multipartite, chen2023efficient, huang2024adaptive}.

Machine learning has emerged as a powerful alternative for addressing complex quantum classification tasks, including entanglement detection \cite{mahdian2025entanglement, mahdian2025optimal, mahdian2025machine, carleo2019machine, sharma2022reformulation, huang2025direct}. Supervised learning techniques, such as support vector machines (SVMs) and neural networks, have been successfully applied to distinguish entangled states from separable ones by learning from labeled training data \cite{lu2018separability, gray2018machine, mahdian2025qsvm, chen2023interpretable}. Unsupervised methods, including clustering and dimensionality reduction, have also been explored for identifying entanglement without prior state knowledge \cite{wang2020unsupervised, koutny2022deep, li2024efficient, abd2025detecting}. However, many ML approaches face challenges in interpretability, computational efficiency, or generalization to unseen states \cite{carleo2019machine, dunjko2018machine, biamonte2017quantum}. In this context, classical  FLDA \cite{fisher1936use, hastie2009elements} offers a compelling balance between simplicity and performance, providing a linear decision boundary that maximizes separability between classes while remaining computationally efficient.

In this work, we leverage FLDA to classify entangled and separable states, combining the interpretability of traditional entanglement criteria with the scalability of machine learning. We establish a theoretical framework for applying FLDA to quantum state discrimination, optimize its implementation for practical use, and rigorously evaluate its performance across different classes of states, including bipartite and multipartite systems. Our results demonstrate that FLDA achieves high classification accuracy with low computational overhead, making it a viable tool for real-world quantum experiments. By bridging classical statistical learning and quantum information, this approach provides a robust and accessible method for entanglement detection, complementing existing techniques while mitigating their limitations.
	
This manuscript is structured as follows: Section 2 introduces the FLDA model, providing a comprehensive overview of its theoretical framework. In Section 3, we employ the FLDA method to classify quantum states comprising two, three, and four qubits, presenting detailed results and analyses. Finally, Section 4 synthesizes the study's findings, offering key conclusions and discussing their implications for future research.

	\section{FLDA}
	FLDA is a classical statistical technique that finds a linear projection to maximize the separation between multiple classes while minimizing the variance within each class. For a dataset with \(n\)-dimensional feature vectors and \(k\) classes, FLDA seeks a projection vector \(\mathbf{W} \in \mathbb{R}^n\) that optimizes the Fisher criterion:

\begin{equation}
	\begin{aligned}
		J(\mathbf{W}) = \frac{\mathbf{W}^T S_B \mathbf{W}}{\mathbf{W}^T S_W \mathbf{W}},
	\end{aligned}
\end{equation}

where \(S_B \in \mathbb{R}^{n \times n}\) is the between-class scatter matrix and \(S_W \in \mathbb{R}^{n \times n}\) is the within-class scatter matrix. These matrices are defined as:

\begin{equation}
	\begin{aligned}
		S_B = \sum_{i=1}^k N_i (\mu_i - \mu)(\mu_i - \mu)^T,
	\end{aligned}
\end{equation}

\begin{equation}
	\begin{aligned}
		S_W = \sum_{i=1}^k \sum_{x \in C_i} (x - \mu_i)(x - \mu_i)^T,
	\end{aligned}
\end{equation}

where \(\mu_i\) is the mean vector of class \(C_i\) and \(\mu\) is the overall mean. \(N_i\) is the number of samples in class \(C_i\), and \(x \in \mathbb{R}^n\) is a feature vector. Maximizing \(J(W)\) ensures that the projected classes are maximally separated.\cite{bishop2006pattern}

To find the optimal \(\mathbf{W} \), we solve the generalized eigenvalue problem:

\begin{equation}
	\begin{aligned}
		S_B \mathbf{W} = \lambda S_W \mathbf{W},
	\end{aligned}
\end{equation}

where the eigenvector \(\mathbf{W} \) corresponding to the largest eigenvalue \(\lambda\) maximizes \(J(\mathbf{W})\). Since the rank of \(S_B\) is at most \(k-1\), FLDA reduces the dimensionality to at most \(k-1\) discriminant directions. If \(S_W\) is singular, which can happen if the number of features is greater than the number of samples, regularization can be applied by adding a small multiple of the identity matrix:
\begin{equation}
	\begin{aligned}
		S_W' = S_W + \epsilon I.
	\end{aligned}
\end{equation}

In discriminant analysis for classification, a new sample $ \mathbf{x} \in \mathbb{R}^n $ is projected into a $ k $-dimensional discriminant subspace defined by the matrix of discriminant vectors $ \mathbf{W} \in \mathbb{R}^{n \times k} $. The projection is computed as $ \mathbf{y} = \mathbf{W}^T \mathbf{x} $, where $ \mathbf{y} \in \mathbb{R}^k $ represents the sample’s coordinates in the reduced-dimensional subspace which has at most $k-1$ dimensions. When $ \mathbf{W} $ is orthonormal, satisfying $ \mathbf{W}^T \mathbf{W} = \mathbf{I}_k $, this mapping is an orthogonal projection, ensuring that $ \mathbf{W} \mathbf{y} $ is the point in the subspace closest to $ \mathbf{x} $ in terms of Euclidean distance. For the special case where $ k = 1 $, the matrix $ \mathbf{W} $ reduces to a single vector $ \mathbf{w} \in \mathbb{R}^n $, and the projection $ y = \mathbf{w}^T \mathbf{x} $ yields a scalar, representing the coordinate of $ \mathbf{x} $ along the line spanned by $ \mathbf{w} $. This dimensionality reduction preserves or enhances class separability, facilitating the use of simplified decision boundaries. Classification is achieved by assigning the sample to the class whose projected mean, $ \mathbf{W}^T \boldsymbol{\mu}_i $, is closest to $ \mathbf{y} $ in the subspace, typically measured using Euclidean distance.

\section{Quantum Entanglement Detection}

In quantum mechanics, the most general density matrix $\rho$ for an $n$-qubit system, residing in the Hilbert space $(\mathbb{C}^2)^{\otimes N}$, is a Hermitian, positive semi-definite operator of dimension $\mathcal{L}((\mathbb{C}^2)^{\otimes N})$ which is the space of all $2^N \times 2^N$ complex matrices and with trace normalized to unity, i.e., $\rho^\dagger = \rho$, $\rho \geq 0$, and $tr(\rho) = 1$. It can be parameterized using the Pauli basis expansion as $\rho = \frac{1}{2^n} \sum_{\mathbf{k}} c_{\mathbf{k}} \sigma^{\mathbf{k}}$, where $\mathbf{k} = (k_1, \dots, k_n)$ with each $k_i \in {0,1,2,3}$, $\sigma^{\mathbf{k}} = \sigma_{k_1} \otimes \cdots \otimes \sigma_{k_n}$ (with $\sigma_0 = I$, $\sigma_1 = X$, $\sigma_2 = Y$, $\sigma_3 = Z$), and the coefficients $c_{\mathbf{k}}$ are real numbers satisfying $|c_{\mathbf{k}}| \leq 1$ for purity constraints, with $c_{\mathbf{0}} = 1$ due to normalization. The parameters $c_{\mathbf{k}}$ (for $\mathbf{k} \neq \mathbf{0}$) form the Bloch vector generalization, totaling $4^n - 1$ independent real parameters, as Hermiticity and positivity impose constraints.
These parameters are specified through measurements via expectation values of observables. Specifically, each $c_{\mathbf{k}} = \langle \sigma^{\mathbf{k}} \rangle = tr(\rho \sigma^{\mathbf{k}})$, obtained by measuring the multi-qubit Pauli operator $\sigma^{\mathbf{k}}$ on an ensemble of identically prepared systems, where the expectation value is the average outcome over multiple trials, leveraging the Born rule for probabilities in projective measurements.

For projective measurements with orthogonal projectors ${\Pi_i}$ (Hermitian, idempotent, mutually orthogonal, and $\sum_i \Pi_i = \mathbb{I}$), the probability of outcome $i$ for density operator $\rho$ is $p_i = tr(\rho \Pi_i)$, normalizing as $\sum_i p_i = 1$. For a Hermitian observable $O = \sum_i o_i \Pi_i$, the expectation value is $\langle O \rangle = tr(\rho O) = \sum_i o_i p_i$, linking probabilities to averages for quantities like spin or energy. 
These measurement outcomes form the basis for classifying quantum states, as the probabilistic nature of the results, combined with the high dimensionality of the Hilbert space, requires transforming quantum data into a classical feature space for analysis, posing challenges for classical classification methods like FLDA.

In quantum systems, feature vectors are built from expectation values of observables \(\{O_i\}\), where \(x_i = \text{tr}(\rho O_i)\). The feature space dimensionality depends on the number of observables.  FLDA reduces this dimensionality, mitigating the curse of dimensionality. The discriminant vector \(\mathbf{W}\) reveals which observables best distinguish quantum states, providing physical insight.

FLDA assumes:1- Normality, Feature vectors per class are approximately normally distributed.2-Homoscedasticity, Classes share equal covariance matrices.3-Linear separability, Classes are separable by linear boundaries.

Though robust to moderate violations, quantum noise and entanglement may challenge homoscedasticity, while the central limit theorem supports normality for averaged measurements.

Feature vectors arise from measurement outcomes like probabilities \( p_i = \text{Tr}(\rho \Pi_i) \) or expectation values \( \langle O \rangle = \text{Tr}(\rho O) \).

The dimensionality $n$ of the full feature space (e.g., the number of all non-identity Pauli operators) scales exponentially with the number of qubits ($4^N - 1$). FLDA reduces this to \( k-1 \), and \(\mathbf{W}\) identifies key measurements (e.g., Pauli operators).

Applying FLDA to quantum data involves:
\begin{itemize}
	\item \textbf{Feature Selection}: Select measurements capturing properties like entanglement efficiently.
	\item \textbf{Noise Handling}: Noise typically increases the within-class scatter $S_W$ and can also distort the between-class scatter $S_B$, requiring filtering or robust statistics.
	\item \textbf{Computational Efficiency}: Computing \( S_B \), \( S_W \), and eigenvalue decomposition (complexity \( O(n^3) \)) may need prior dimensionality reduction.
\end{itemize}
We examine Positive Partial Transpose (PPT) states, which are bipartite or multipartite mixed quantum states whose density matrix remains positive semi-definite after partial transposition over one subsystem (e.g., $\rho^{T_A}$). These states, characterized by tunable entanglement, are ideal for classification tests in quantum information theory. For low-dimensional systems (e.g., $2\times2$ or $2\times3$), the Peres-Horodecki criterion ensures PPT states are separable, while in higher dimensions, some entangled states may be PPT, known as bound entangled states. We describe two-, three-, and four-qubit Werner states, with a focus on three-qubit PPT states as classified in \cite{acin2001classification}, highlighting their role in entanglement analysis.

FLDA offers an efficient framework for quantum state classification by mapping measurements to a classical feature space. Its sensitivity to non-linear separability and noise requires careful feature engineering for optimal performance.

\subsection{Two-Qubit }
\label{sec:two_qubit}

A two-qubit Werner state in \(\mathbb{C}^2 \otimes \mathbb{C}^2 \) interpolates between a maximally entangled state and the maximally mixed state. For a general Bell state \(|\Psi\rangle\), it is expressed in the Pauli basis as:

$c_{ij} = 0$ for $i \neq j$, and for the diagonal terms $ c_{ii}$ we have

\begin{equation}
	\rho_{Wer}^{(2)} = \frac{1}{4} \sum_{i,j=0}^{3} c_{ij} (\sigma_i \otimes \sigma_j), \quad  c_{ii} =   
	\begin{cases}
		1 & \text{if } i = 0, \\
		p s_i & \text{if } i = 1,2,3.
	\end{cases}
\end{equation}
where \(p \in [-\frac{1}{3},1]\) is the mixing parameter and $s_i = \pm 1$ depends on the Bell state.

The entanglement of the two-qubit Werner state \(\rho_{\text{Wer}}^{(2)}\) is assessed using the Peres-Horodecki criterion. The critical eigenvalue condition \(\frac{1-3p}{4} < 0\) implies \(p > \frac{1}{3}\). Thus, the state is separable for \(p \leq \frac{1}{3}\) and entangled for \(p > \frac{1}{3}\), where the partial transpose yields a negative eigenvalue.

\begin{figure}[H]
	\centering
	\includegraphics[width=0.6\linewidth]{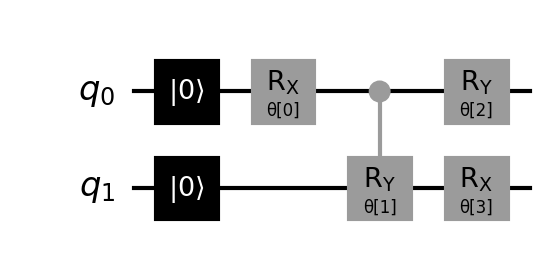}
	\caption{The  final state is expressed as \( |\psi\rangle = \cos\left(\frac{\theta_0}{2}\right) |00\rangle - i \sin\left(\frac{\theta_0}{2}\right) \cos\left(\frac{\theta_1}{2}\right) |10\rangle - i \sin\left(\frac{\theta_0}{2}\right) \sin\left(\frac{\theta_1}{2}\right) |11\rangle \). For \( \theta_0 = \frac{\pi}{2}, \theta_1 = \pi \), the state is maximally entangled (\( C = 1 \)); for \( \theta_0 = 0, \theta_1 = \pi \), it is separable (\( C = 0 \)). \cite{mahdian2025entanglement}}
	\label{fig:2_qubit}
\end{figure}

To prepare a two-qubit Werner state, the singlet state \( |\Psi^-\rangle \) is generated using a quantum circuit:
$
|\Psi^-\rangle = \text{CNOT} \cdot (H \otimes I) \cdot |01\rangle,
$
followed by a depolarizing channel
$
\mathcal{E}(\rho) = p \rho + \frac{1-p}{4} I_4.
$
To quantify the entanglement, we employ the concurrence method, where the concurrence \( C \) is computed as \( C = \sin(\theta_0) \sin\left(\frac{\theta_1}{2}\right) \) form quantum circuit (see Fig.~\ref{fig:2_qubit}). Here, $\sin(\theta_0)$ reflects the control qubit’s initial superposition amplitude . A concurrence of 0 indicates a separable state, while a value of 1 confirms a maximally entangled state. 

\begin{figure}[H]
	\centering
	\begin{subfigure}[b]{0.9\linewidth}
		\centering
		\includegraphics[width=\linewidth]{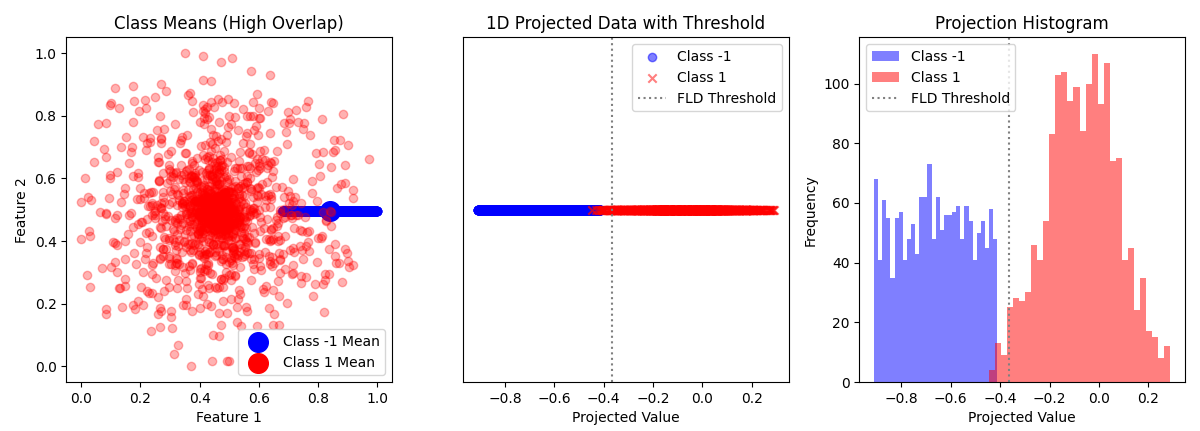}
		\caption{High Overlap}
		\label{fig:ldaqda_high}
	\end{subfigure}
	\hfill
	\begin{subfigure}[b]{0.9\linewidth}
		\centering
		\includegraphics[width=\linewidth]{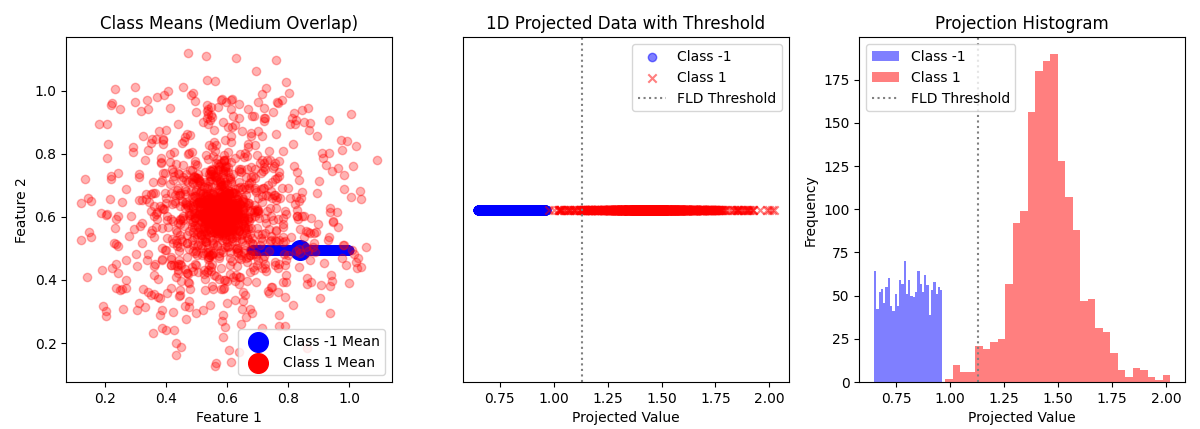}
		\caption{Medium Overlap}
		\label{fig:ldaqda_medium}
	\end{subfigure}
	\hfill
	\begin{subfigure}[b]{0.9\linewidth}
		\centering
		\includegraphics[width=\linewidth]{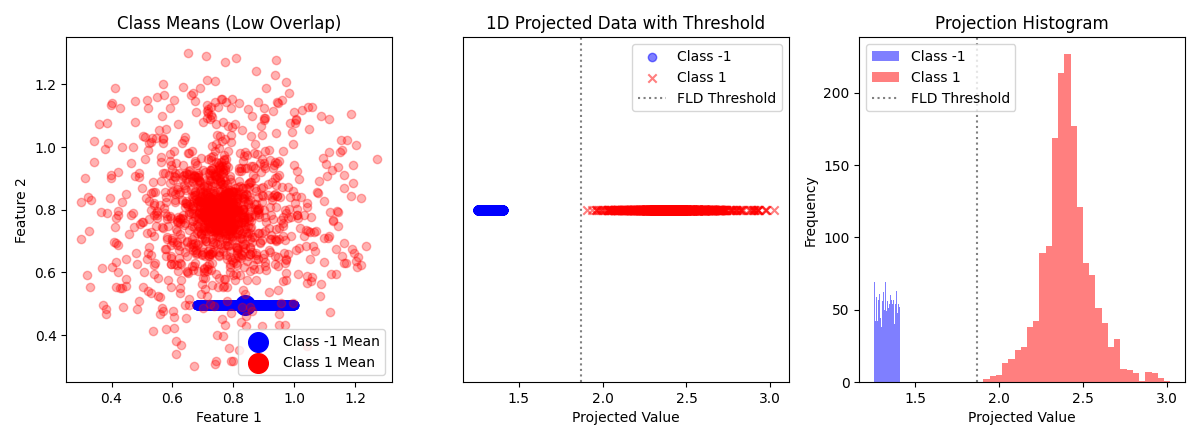}
		\caption{Low Overlap}
		\label{fig:ldaqda_low}
	\end{subfigure}
	\caption{Comparison of FLDA plots with different overlaps two-qubit entangled Werner states and product separable states.  Class -1 indicates the entangled states, whereas Class 1 signifies the separable states.}
	\label{fig:lda_comparison2}
\end{figure}

\begin{figure}[H]
	\centering
	\begin{subfigure}[b]{0.9\linewidth}
		\centering
		\includegraphics[width=\linewidth]{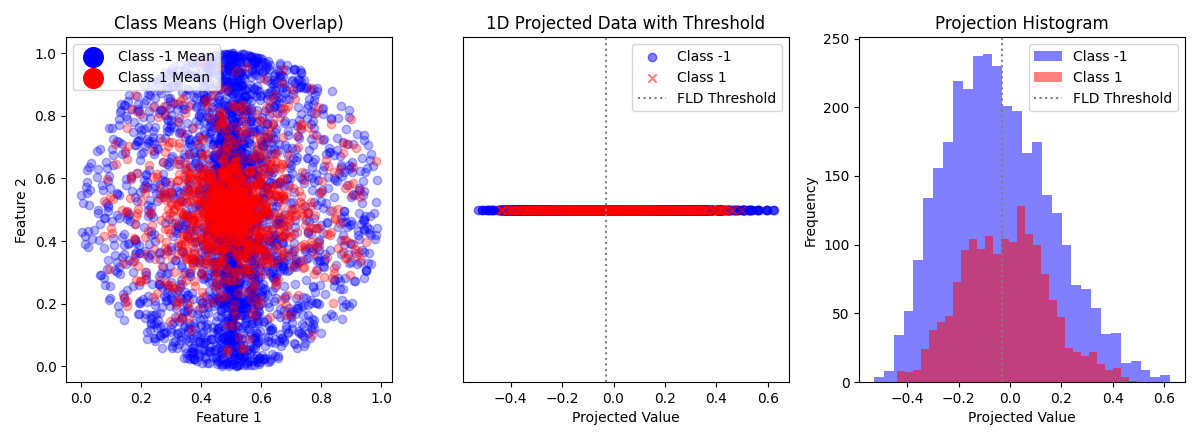}
		\caption{High Overlap}
		\label{fig:ldaqda_high}
	\end{subfigure}
	\hfill
	\begin{subfigure}[b]{0.9\linewidth}
		\centering
		\includegraphics[width=\linewidth]{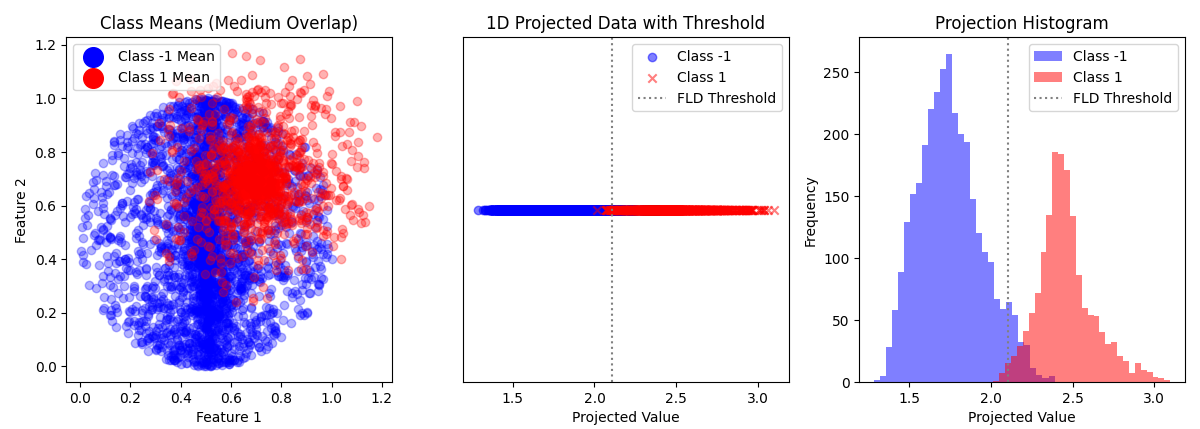}
		\caption{Medium Overlap}
		\label{fig:ldaqda_medium}
	\end{subfigure}
	\hfill
	\begin{subfigure}[b]{0.9\linewidth}
		\centering
		\includegraphics[width=\linewidth]{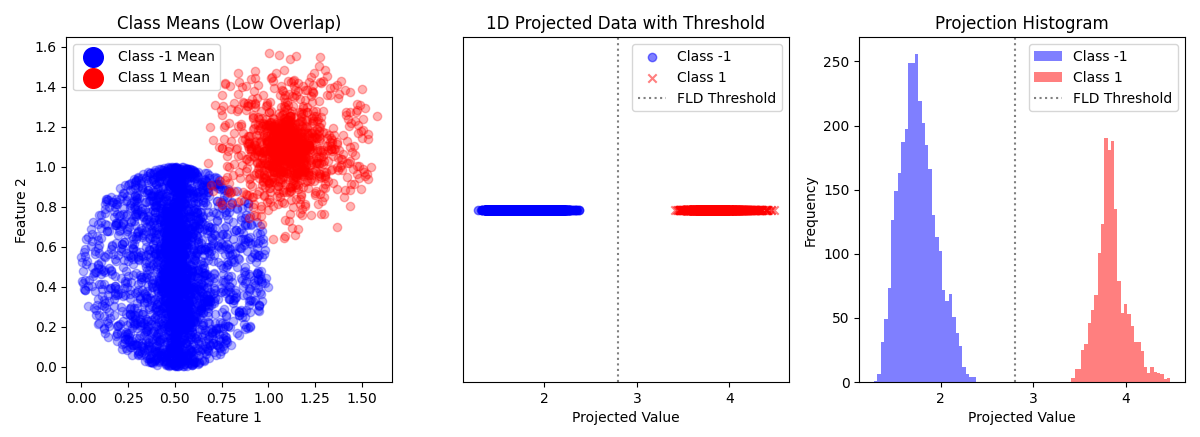}
		\caption{Low Overlap}
		\label{fig:ldaqda_low}
	\end{subfigure}
	\caption{Comparison of FLDA plots with different overlaps two-qubit entangled with concurrence method and product separable states. Class -1 indicates the entangled states, whereas Class 1 signifies the separable states.}
	\label{fig:lda_comparison1}
\end{figure}

A two-qubit density matrix is prepared using Werner states or concurrence, then measured to extract classical data for FLDA. Observables like \(\sigma_x, \sigma_y, \sigma_z\) and correlations \(\sigma_i \otimes \sigma_i\) are measured over multiple trials (e.g., 1000) to estimate expectation values (e.g., \(\langle \sigma_z \otimes \sigma_z \rangle\)) and probabilities. These form feature vectors \(x \in \mathbb{R}^n\), labeled as separable (\(p \leq \frac{1}{3}\)) or entangled (\(p > \frac{1}{3}\)), creating a dataset of \(N=10000\) samples (e.g., 5000 per class). Features are normalized to [0, 1] or standardized, and \(S_W\) is regularized (\(S_W' = S_W + \epsilon I\)) if needed. FLDA computes class means \(\mu_1, \mu_2\), matrices \(S_B, S_W\), and solves \(S_B w = \lambda S_W w\) for the projection vector \(w\). For two classes, FLDA projects onto a scalar \(y = w^T x\), with a threshold (e.g., midpoint of \(w^T \mu_1, w^T \mu_2\)) for classification (see Fig.~\ref{fig:lda_comparison1}, \ref{fig:lda_comparison2}), as applied to Werner states  (Eq.\ref{eq:w_werner}) and concurrence-quantified states  (Eq.\ref{eq:w_concurace}).

\begin{equation}
	W_{Wer}=  
	\begin{bmatrix}
		0.12 & 0.368 & 0.204 & 0.170 & 0.455 \\
		0.652 &0.188& 0.151& 0.187 & 0.204   \\
		0.033 &0.059 &	0.106 & 0.098 & 0.051
	\end{bmatrix},
	\label{eq:w_werner}
\end{equation}

\begin{equation}
	W_{Con}=  
	\begin{bmatrix}
		0.501 & 0.244 & 0.335 & 0.183 & 0.311 \\
		0.391 &0.242& 0.157& 0.185 & 0.295   \\
		0.159 &0.099 &	0.132 & 0.160 & 0.092
	\end{bmatrix}.
	\label{eq:w_concurace}
\end{equation}

$W$ defined in both Eq.\ref{eq:w_werner},Eq.\ref{eq:w_concurace} is a $1 \times 15$ projection vector.
Considering a new Werner state with unknown $p$, the same observables are measured, the feature vector $x$ is constructed, and projected to $\mathbf{y}=\mathbf{W}^{T}\mathbf{x}$. The state is classified as separable if $\mathbf{y}$ is closer to $W^T \mu_1$, or entangled if closer to $W^T \mu_2$.

The accuracy of the model is validated using a test set, with a particular focus on the region near the boundary $p = \frac{1}{3}$, where noise and measurement uncertainties pose significant challenges Tab.\ref{tab:overlap_results_wer}. The results for the concurrence method are presented in Tab.\ref{tab:overlap_results_con}.

	\begin{algorithm}
	\caption{Two-Qubit Entanglement Detection via FLDA}
	\label{alg:two_qubit_entanglement}
	\begin{algorithmic}[1]
		\Require Quantum state parameters: $p$ (Werner) or $\theta_0, \theta_1$ (concurrence)
		\Require Observables: $\{\sigma_x, \sigma_y, \sigma_z, \sigma_i \otimes \sigma_i\}$ for $i = 1,2,3$
		\Require $N_{\text{trials}} = 1000$, $N_{\text{train}} = 10000$ (5000 per class)
		\Ensure Classification: ``Entangled'' or ``Separable''
		
		\Statex
		\Function{Detect Entanglement}{}
		\State Prepare two-qubit quantum state
		\If{using Werner state}
		\State $\rho_{\text{Wer}}^{(2)} \gets \frac{1}{4} \sum_{i,j=0}^3 c_{ij} (\sigma_i \otimes \sigma_j)$
		\State $c_{ij} \gets \delta_{ij} \cdot \begin{cases} 1 & \text{if } i=0 \\ p s_i & \text{if } i=1,2,3 \end{cases}$
		\Else
		\State $|\psi\rangle \gets \cos\left(\frac{\theta_0}{2}\right)|00\rangle - i\sin\left(\frac{\theta_0}{2}\right)\cos\left(\frac{\theta_1}{2}\right)|10\rangle - i\sin\left(\frac{\theta_0}{2}\right)\sin\left(\frac{\theta_1}{2}\right)|11\rangle$
		\EndIf
		\For{each observable $O$}
		\State Perform $N_{\text{trials}}$ measurements
		\State $\langle O \rangle \gets \operatorname{tr}(\rho O)$ 
		 {Estimate expectation value}
		\EndFor
		\State $\mathbf{x} \gets [\langle O_1 \rangle, \langle O_2 \rangle, \dots, \langle O_n \rangle]^T$ {Construct feature vector}
		\State Normalize/standardize features
		
		\If{$p \leq \frac{1}{3}$ (Werner) or $C = 0$ (concurrence)}
		\State $\text{label} \gets \text{``Separable''}$
		\Else
		\State $\text{label} \gets \text{``Entangled''}$
		\EndIf
		
		\State Compute $S_W$ and $S_B$ matrices:
		\State $S_B \gets \sum_{i=1}^k N_i (\boldsymbol{\mu}_i - \boldsymbol{\mu})(\boldsymbol{\mu}_i - \boldsymbol{\mu})^T$
		\State $S_W \gets \sum_{i=1}^k \sum_{\mathbf{x} \in C_i} (\mathbf{x} - \boldsymbol{\mu}_i)(\mathbf{x} - \boldsymbol{\mu}_i)^T$
		
		\State Solve $S_B \mathbf{w} = \lambda S_W \mathbf{w}$ {Generalized eigenvalue problem}
		\State Obtain projection vector $\mathbf{w}$
		
		\For{new state $\mathbf{x}_{\text{new}}$}
		\State $y \gets \mathbf{w}^T \mathbf{x}_{\text{new}}$ {Project onto discriminant direction}
		\If{$y$ closer to $\mathbf{w}^T \boldsymbol{\mu}_{\text{entangled}}$}
		\State \Return ``Entangled''
		\Else
		\State \Return ``Separable''
		\EndIf
		\EndFor
		\EndFunction
	\end{algorithmic}
\end{algorithm}

\begin{table}[htbp]
	\centering
	\scriptsize
	\begin{tabular}{|l|c|c|c|c|}
		\hline
		Overlap Level & FLD Threshold & Train Accuracy & Test Accuracy & Fisher Criterion \\
		\hline
		High & -0.36 & 0.94 & 0.89 & 8.41 \\
		\hline
		Medium & 1.13 & 0.95 & 0.92 & 13.65 \\
		\hline
		Low & 1.86 & 100 & 1.00 & 45.43 \\
		\hline
	\end{tabular}
	\caption{Summary of classification performance for different state overlaps with quantum measurements from Werner states.}
	\label{tab:overlap_results_wer}
\end{table}

\begin{table}[htbp]
	\centering
	\scriptsize
	\begin{tabular}{|l|c|c|c|c|}
		\hline
		Overlap Level & FLD Threshold & Train Accuracy & Test Accuracy & Fisher Criterion \\
		\hline
		High & 0.41 & 0.66 & 0.33 & 0.012 \\
		\hline
		Medium & 2.09 & 0.95 & 0.83 & 7.43 \\
		\hline
		Low & 2.23 & 1.00 & 1.00 & 44.6 \\
		\hline
	\end{tabular}
	\caption{Summary of classification performance for different state overlaps for concurrence method.}
	\label{tab:overlap_results_con}
\end{table}

\vspace{1.5cm}

This approach leverages measurement outcomes to bridge quantum states and classical FLDA, with features for two-qubit Werner states, and extends to three- and four-qubit states, where additional observables and classifications enrich the feature space despite the growing complexity of the Hilbert space.

\subsection{Three-Qubit Werner States}

For three qubits, the classification of mixed three-qubit states, as detailed in \cite{acin2001classification}, involves categorizing states based on their separability across different bipartitions using the PPT criterion. 
A three-qubit Werner state, based on a general entangled state \(|\Psi_{GHZ}\rangle\), interpolates between a maximally entangled state and the maximally mixed state. 
$c_{ijk} = 0$ for $i \neq j \neq k $, and for the diagonal terms $ c_{iii}$ we have
\begin{equation}
	\rho_{Wer}^{(3)} = \frac{1}{8}  \sum_{i,j,k \in \{0,1,2,3\}} c_{ijk} (\sigma_i \otimes \sigma_j \otimes \sigma_k) , c_{iii} = 
	\begin{cases}
		1 & \text{if } i = 0, \\
		p \, s_i & \text{if } i = 1,2,3.
	\end{cases}
\end{equation}
\(s_{i} = \pm 1\) are state-dependent coefficients. The PPT criterion for the three-qubit Werner state yields a critical eigenvalue \(\frac{1-3p}{8} \geq 0\), indicating full separability for \(p \leq \frac{1}{3}\). The state is separable across all bipartitions for \(p \leq \frac{1}{5}\), but typically exhibits genuine tripartite entanglement for \(p > \frac{1}{5}\), with the PPT threshold at \(p > \frac{1}{3}\).

A tripartite quantum state is \emph{fully separable} if and only if it admits a decomposition of the form
$
\rho_{\text{sep}} = \sum_i p_i \, \rho_A^i \otimes \rho_B^i \otimes \rho_C^i.$
where $\{p_i\}$ form a probability distribution and $\rho_X^i$ denote quantum states for subsystem $X$. Such states represent the absence of entanglement across all possible bipartitions (A|BC, B|AC, C|AB) and can be prepared through local operations and classical communication (LOCC).
A more general class of \emph{biseparable states} exhibits separability only with respect to specific bipartitions. For example, a state separable under the A|BC partition but potentially entangled across other cuts can be expressed as
\[
\rho_{\text{bisep}} = \sum_i p_i \, \rho_A^i \otimes \rho_{BC}^i.
\]
where $\rho_{BC}^i$ may contain entanglement between subsystems B and C. The existence of such states demonstrates that partial separability does not imply full separability in multipartite systems.

Positive partial transpose (PPT) entangled states are quantum states that remain positive semi-definite under partial transposition, a key criterion in quantum information theory for assessing entanglement. Unlike separable states, PPT entangled states exhibit non-trivial quantum correlations, yet their partial transpose has non-negative eigenvalues, distinguishing them from states that violate the PPT criterion (NPT states). These states are significant in quantum information processing, as they often represent bound entanglement, which cannot be distilled into pure entangled states. Below, we present two representations of PPT entangled states, as described in the literature. \cite{guhne2009entanglement}

\begin{itemize}
	\item \textbf{Generalized PPT Entangled State}

The density matrix for a class of PPT entangled states in a three-qubit system, as introduced by \cite{acin2001classification}, is given by:

\begin{equation}
	\label{eq:pptes}
	\rho_{\text{PPTES}} = \frac{1}{n} \begin{bmatrix}
		1 & 0 & 0 & 0 & 0 & 0 & 0 & 1 \\
		0 & a & 0 & 0 & 0 & 0 & 0 & 0 \\
		0 & 0 & b & 0 & 0 & 0 & 0 & 0 \\
		0 & 0 & 0 & c & 0 & 0 & 0 & 0 \\
		0 & 0 & 0 & 0 & \frac{1}{c} & 0 & 0 & 0 \\
		0 & 0 & 0 & 0 & 0 & \frac{1}{b} & 0 & 0 \\
		0 & 0 & 0 & 0 & 0 & 0 & \frac{1}{a} & 0 \\
		1 & 0 & 0 & 0 & 0 & 0 & 0 & 1 \\
	\end{bmatrix},
\end{equation}
where the normalization constant is defined as:
\[
n = 2 + a + \frac{1}{a} + b + \frac{1}{b} + c + \frac{1}{c}.
\]
and the parameters satisfy $a, b, c > 0$. This matrix represents a valid density operator, being Hermitian, positive semi-definite, and having a trace of 1. The non-zero off-diagonal elements (1 in the top-right and bottom-left corners) indicate quantum correlations, while the positive parameters $a$, $b$, and $c$ ensure the PPT property when the partial transpose is computed with respect to one or more subsystems.
\end{itemize}

\begin{itemize}
\item \textbf{Alternative Representation of PPT States}

Another form of a PPT entangled state for a three-qubit system can be expressed using Pauli operators, as noted in the referenced article:

\begin{equation}
	\label{eq:pptes_pauli}
	\rho = \frac{1}{8} \left( I \otimes I \otimes I + I \otimes \sigma_z \otimes \sigma_z + \sigma_z \otimes I \otimes \sigma_z + \sigma_z \otimes \sigma_z \otimes I \right).
\end{equation}
where $I$ is the $2 \times 2$ identity matrix and $\sigma_z$ is the Pauli Z matrix.
This representation leverages the tensor product structure of the three-qubit Hilbert space. The combination of identity and Pauli $\sigma_z$ operators introduces specific correlations among the qubits, resulting in a density matrix that is PPT with respect to all possible bipartitions yet exhibits entanglement. The factor $\frac{1}{8}$ ensures proper normalization, as the trace of the operator must equal 1 for a valid density matrix.\cite{jafarizadeh2008detecting} 
\end{itemize}

For classification with FLDA, measurements are chosen to capture entanglement properties.
 We apply the same methodology described in Sec.\ref{sec:two_qubit} for  \( \rho_{Wer}^{(3)} \) and 
 all the three-qubits. 

\begin{figure}[H]
	\centering
	\begin{subfigure}[b]{0.9\linewidth}
		\centering
		\includegraphics[width=\linewidth]{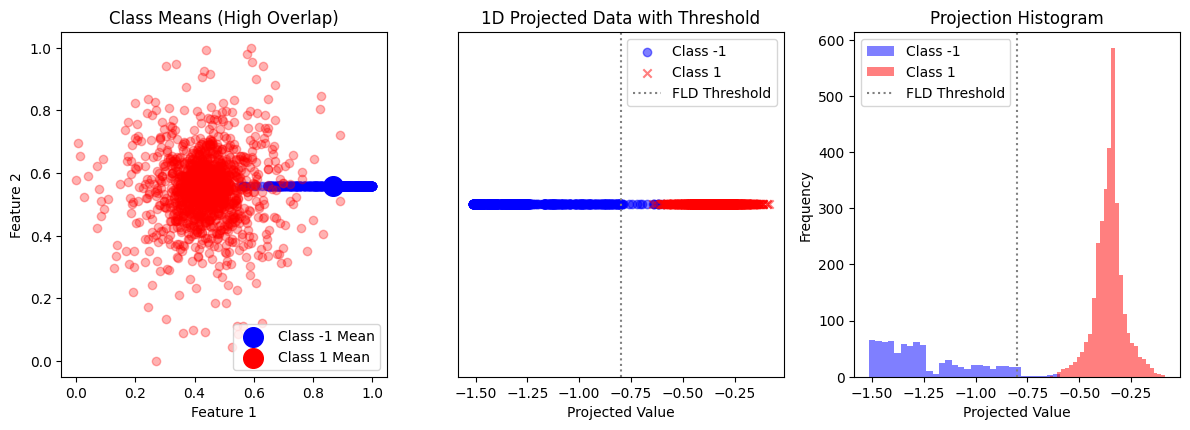}
		\caption{High Overlap}
		\label{fig:ldaqda_high}
	\end{subfigure}
	\hfill
	\begin{subfigure}[b]{0.9\linewidth}
		\centering
		\includegraphics[width=\linewidth]{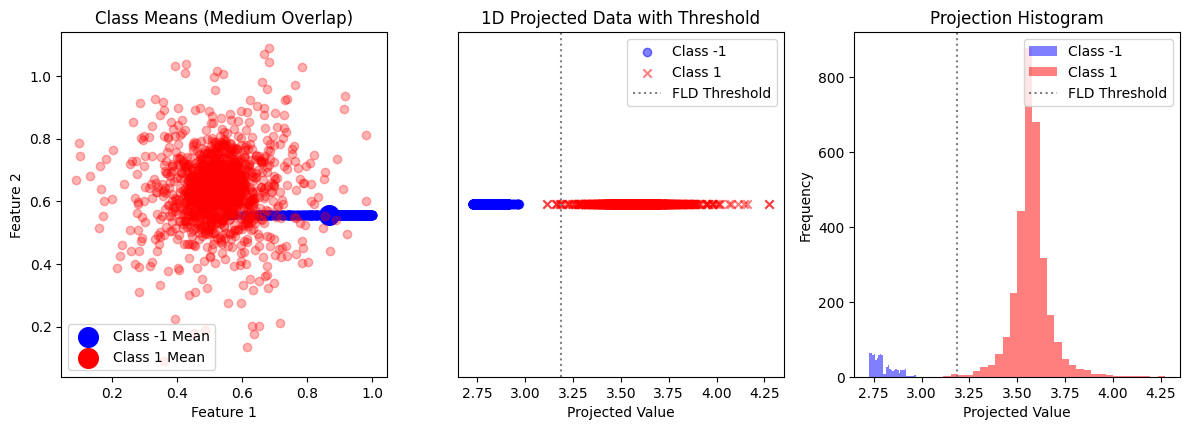}
		\caption{Medium Overlap}
		\label{fig:ldaqda_medium}
	\end{subfigure}
	\hfill
	\begin{subfigure}[b]{0.9\linewidth}
		\centering
		\includegraphics[width=\linewidth]{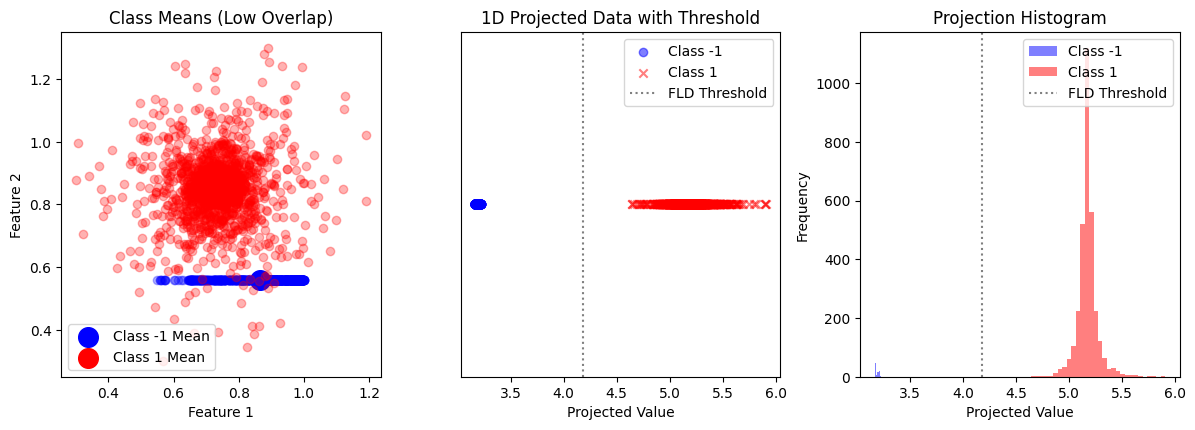}
		\caption{Low Overlap}
		\label{fig:ldaqda_low}
	\end{subfigure}
	\caption{Comparison of FLDA plots with different overlaps three-qubit GHZ-based Werner States and product separable states. Class -1 indicates the entangled states, whereas Class 1 signifies the separable states.}
	\label{fig:lda_comparison1_3qubit}
\end{figure}

\begin{figure}[H]
	\centering
	\begin{subfigure}[b]{0.9\linewidth}
		\centering
		\includegraphics[width=\linewidth]{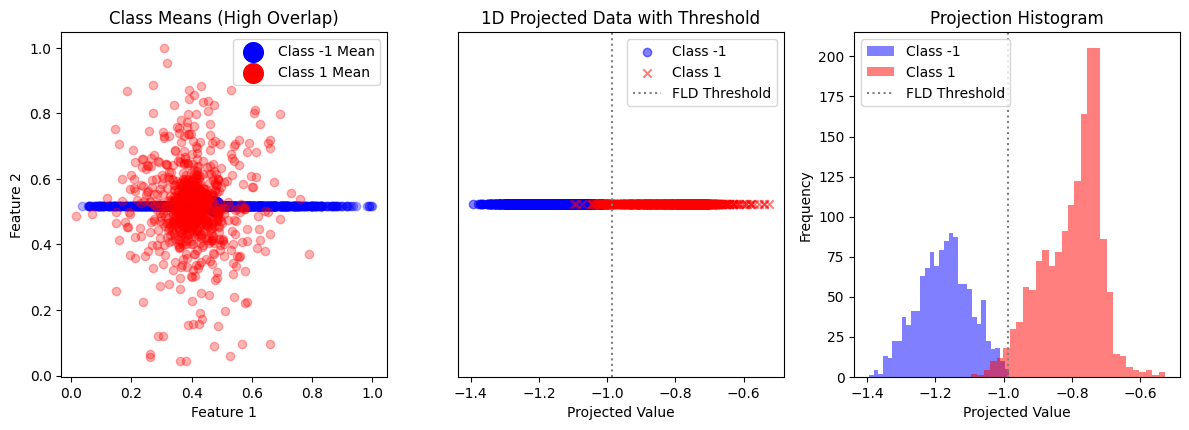}
		\caption{High Overlap}
		\label{fig:ldaqda_high}
	\end{subfigure}
	\hfill
	\begin{subfigure}[b]{0.9\linewidth}
		\centering
		\includegraphics[width=\linewidth]{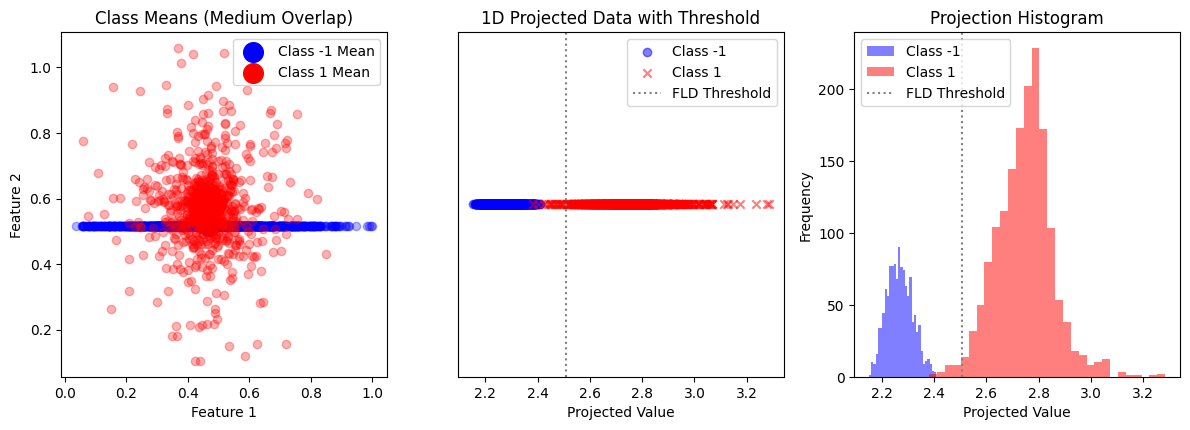}
		\caption{Medium Overlap}
		\label{fig:ldaqda_medium}
	\end{subfigure}
	\hfill
	\begin{subfigure}[b]{0.9\linewidth}
		\centering
		\includegraphics[width=\linewidth]{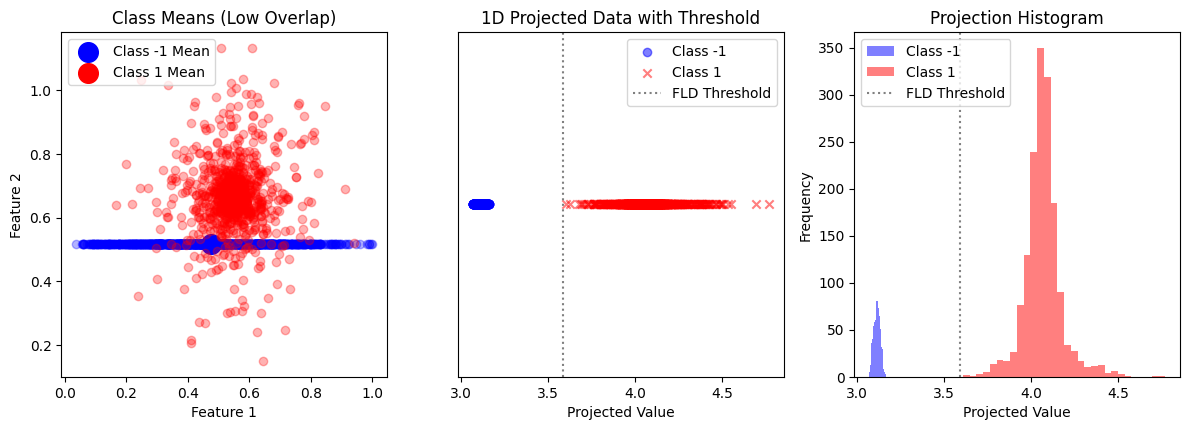}
		\caption{Low Overlap}
		\label{fig:ldaqda_low}
	\end{subfigure}
	\caption{Comparison of FLDA plots with different overlaps three-qubit alternative PPT States and product separable states. Class -1 indicates the entangled states, whereas Class 1 signifies the separable states.}
	\label{fig:lda_comparison1_ppt_a}
\end{figure}

\begin{figure}[H]
	\centering
	\begin{subfigure}[b]{0.9\linewidth}
		\centering
		\includegraphics[width=\linewidth]{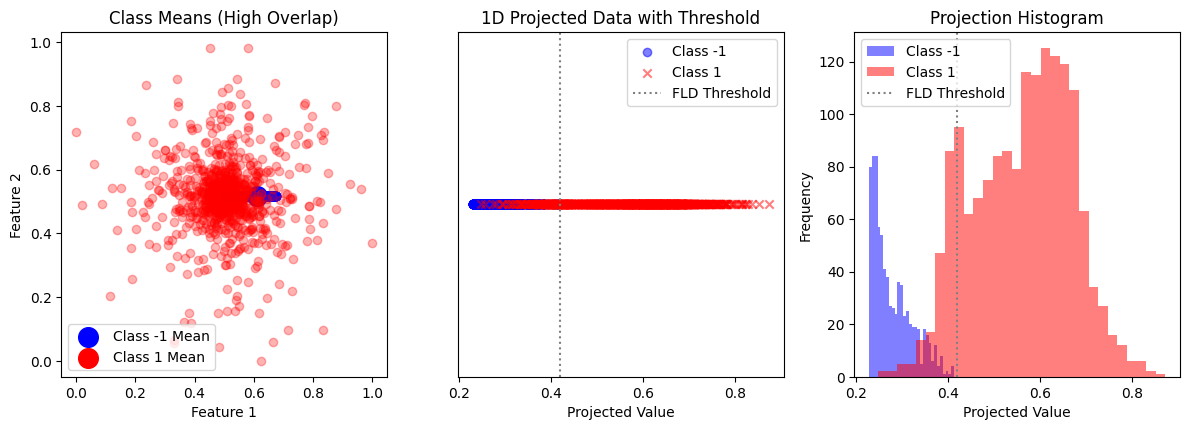}
		\caption{High Overlap}
		\label{fig:ldaqda_high}
	\end{subfigure}
	\hfill
	\begin{subfigure}[b]{0.9\linewidth}
		\centering
		\includegraphics[width=\linewidth]{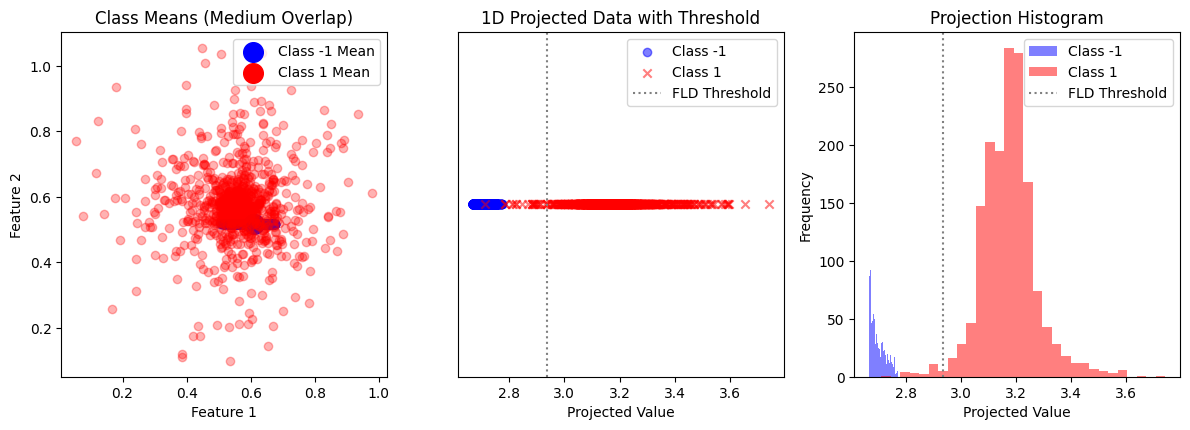}
		\caption{Medium Overlap}
		\label{fig:ldaqda_medium}
	\end{subfigure}
	\hfill
	\begin{subfigure}[b]{0.9\linewidth}
		\centering
		\includegraphics[width=\linewidth]{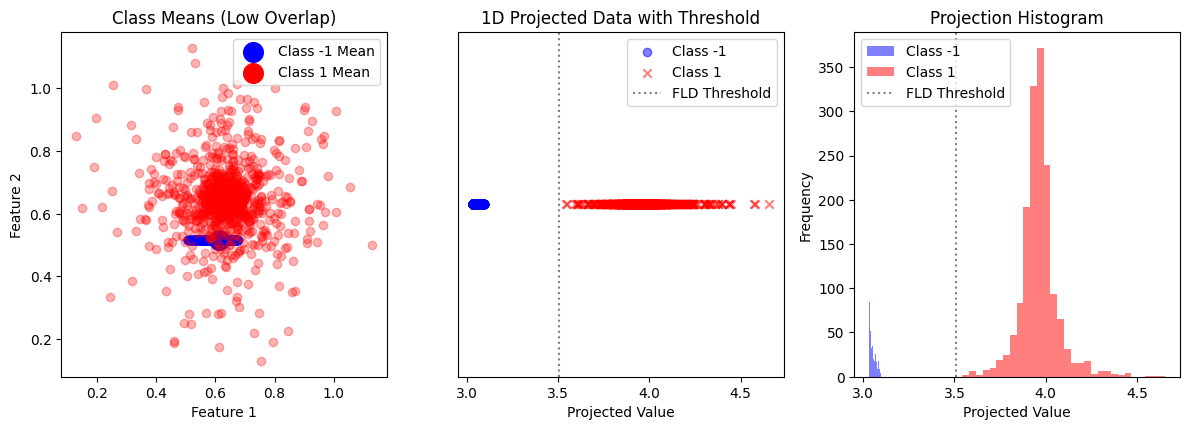}
		\caption{Low Overlap}
		\label{fig:ldaqda_low}
	\end{subfigure}
	\caption{Comparison of FLDA plots with different overlaps three-qubit PPT States and product separable states. Class -1 indicates the entangled states, whereas Class 1 signifies the separable states.}
	\label{fig:lda_comparison1_ppt}
\end{figure}

\begin{figure}
	\centering
	\includegraphics[width=0.7\linewidth]{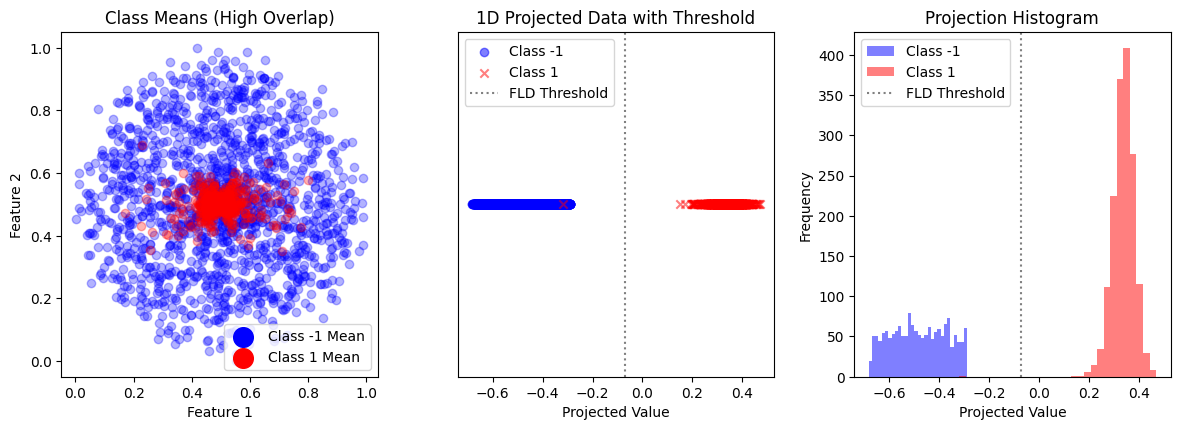}
	\caption{FLDA plots with high overlap for three-qubit biseparable States and product separable states. Class -1 indicates the entangled states, whereas Class 1 signifies the separable states.}
	\label{fig:biparth}
\end{figure}

\begin{table}[htbp]
	\centering
	\scriptsize
	\begin{tabular}{|l|c|c|c|c|}
		\hline
		Overlap Level & FLD Threshold & Train Accuracy & Test Accuracy & Fisher Criterion \\
		\hline
		High & -0.79 & 0.93 & 0.87 & 14.99\\
		\hline
		Medium & 3.18 & 0.98 & 0.96 &  44.24 \\
		\hline
		Low & 4.17 & 1.00 & 1.00 & 353.08 \\
		\hline
	\end{tabular}
	\caption{Summary of classification performance for different state overlaps with quantum measurements for three-qubit GHZ states.}
	\label{tab:overlap_results_wer_3}
\end{table}

\begin{table}[htbp]
	\centering
	\scriptsize
	\begin{tabular}{|l|c|c|c|c|}
		\hline
		Overlap Level & FLD Threshold & Train Accuracy & Test Accuracy & Fisher Criterion \\
		\hline
		High & 0.37 & 0.76 & 0.65 & 5.4\\
		\hline
		Medium & 2.9 & 0.93 & 0.88 &  18.68 \\
		\hline
		Low & 3.56 & 1.00 & 1.00 & 70.69 \\
		\hline
	\end{tabular}
	\caption{Summary of classification performance for different state overlaps with quantum measurements for three-qubit PPT states.}
	\label{tab:overlap_results_ppt}
\end{table}

\begin{table}[H]
	\centering
	\scriptsize
	\begin{tabular}{|l|c|c|c|c|}
		\hline
		Overlap Level & FLD Threshold & Train Accuracy & Test Accuracy & Fisher Criterion \\
		\hline
		High & -0.98 & 0.90 & 0.82 & 9.48\\
		\hline
		Medium & 3.18 & 0.89 & 0.80 &  15.2 \\
		\hline
		Low & 4.17 & 1.00 & 1.00 & 69.51 \\
		\hline
	\end{tabular}
	\caption{Summary of classification performance for different state overlaps with quantum measurements for three-qubit alternative PPT states.}
	\label{tab:overlap_results_ppt_a}
\end{table}

\begin{table}[H]
	\centering
	\scriptsize
	\begin{tabular}{|l|c|c|c|c|}
		\hline
		Overlap Level & FLD Threshold & Train Accuracy & Test Accuracy & Fisher Criterion \\
		\hline
		High & -0.07 & 1.00 & 1.00 & 45.23\\
		\hline

	\end{tabular}
	\caption{Summary of classification performance for high overlaps with quantum measurements for three-qubit bipartite states.}
	\label{tab:overlap_results_bipart}
\end{table}

\subsection{Four-Qubit Werner States}

For four-qubit systems, the classification of mixed states extends naturally from the tripartite case \cite{acin2001classification}, with separability properties analyzed across all possible bipartitions (e.g., A|BCD, AB|CD, etc.). The four-qubit Werner state constructed from a generalized GHZ state, interpolates between maximal entanglement and the maximally mixed state.
the 4-qubit GHZ state in the Pauli basis representation: 
\begin{equation}
	\rho_{\text{Wer}}^{(4)} = \frac{1}{16} \sum_{\substack{i,j,k,l \in \{0,1,2,3\}}} c_{ijkl} (\sigma_i \otimes \sigma_j \otimes \sigma_k \otimes \sigma_l) .
\end{equation}
Entanglement is detected for $p > \frac{1}{7}$, with separability for  $p\leq \frac{1}{7}$. This threshold reflects the increased bipartition complexity in four-qubit systems compared to the tripartite case.

Preparation follows a similar circuit-based approach, with measurements focusing on multi-qubit correlations.
The high dimensionality of the feature space for multi-qubit systems and the presence of measurement noise can increase the within-class scatter \( S_W \), reducing FLDA’s effectiveness. Careful selection of measurements is crucial to ensure linear separability.
The discriminant vector \( W \) from FLDA indicates which measurements are most informative, potentially guiding experimental design by identifying key observables for entanglement detection.

\begin{figure}[H]
	\centering
	\includegraphics[width=0.7\linewidth]{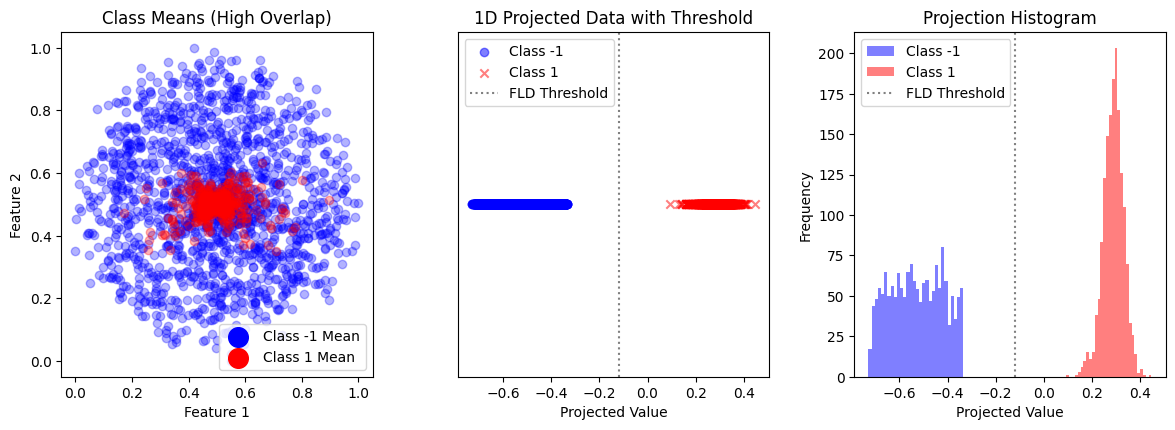}
	\caption{FLDA plots with high overlap for four-qubit Werner State and product separable states. Class -1 indicates the entangled states, whereas Class 1 signifies the separable states.}
	\label{fig:4qubith}
\end{figure}

\begin{table}[H]
	\centering
	\scriptsize
	\begin{tabular}{|l|c|c|c|c|}
		\hline
		Overlap Level & FLD Threshold & Train Accuracy & Test Accuracy & Fisher Criterion \\
		\hline
		High & -0.12 & 1.00 & 1.00 & 45.99\\
		\hline
		
	\end{tabular}
	\caption{Summary of classification performance for high overlaps with quantum measurements for four-qubit Werner state.}
	\label{tab:overlap_results_4qubit}
\end{table}

	\section{Discussion and Conclusion}
	Our results demonstrate that FLDA can effectively classify quantum states when measurement statistics are sufficiently distinct. The performance naturally degrades as the classes of states become less distinguishable (i.e., have higher overlap), as shown by the decreasing test accuracy and Fisher criterion values. However, even in high-overlap scenarios, FLDA often maintains high classification accuracy (e.g., 89\% for two-qubit and 87\% for three-qubit Werner states).
	
	The tests on PPT entangled and biseparable states highlight the method's behavior in more complex scenarios. The lower accuracy for the PPT entangled state suggests that its entanglement is not strongly captured by the linear combination of Pauli observables found by FLDA. In contrast, the perfect accuracy for the biseparable state in our noise-free simulation indicates that its structure is easily distinguished from fully separable states by the chosen measurements. It is important to note that these results are from idealized simulations. The presence of experimental noise would inflate the within-class scatter (\(S_W\)) and almost certainly lower these accuracy figures, making the classification more challenging in practice.
	
	A key advantage of FLDA is its interpretability. The discriminant vector  \(W\) reveals which observables are most influential for classification, which can guide the design of more efficient measurement strategies by focusing on the most informative observables. This is particularly valuable for applications like quantum state tomography and quantum error correction.
	
	Overall, this study demonstrates that FLDA is a simple, efficient, and interpretable framework for quantum state classification. By transforming quantum measurement outcomes into a classical feature space, FLDA effectively leverages statistical techniques to distinguish between different classes of quantum states, including separable, entangled, and biseparable states. Our results, based on clear and reproducible simulations, show that the method achieves high classification accuracy and remains effective even in challenging high-overlap scenarios.
	
	The interpretability of the FLDA discriminant provides valuable physical insight, identifying the most informative measurements for distinguishing quantum states, which can help optimize experimental designs. The successful application to two-, three-, and four-qubit systems highlights the method's versatility. However, FLDA relies on the assumption of linear separability in the chosen feature space, which may not hold for all quantum systems. Performance is also sensitive to measurement noise, which was not included in our idealized simulations but would be a critical factor in any real-world application. The high dimensionality of multi-qubit systems also necessitates a large number of measurements to define the feature space, although FLDA itself provides a robust mechanism for dimensionality reduction.
	
	In conclusion, FLDA represents a promising and practical tool for quantum state classification, balancing simplicity, efficiency, and interpretability. It provides a strong baseline for classification tasks and a valuable guide for measurement optimization. Future work should focus on integrating this method with more advanced feature selection techniques and testing its robustness against realistic experimental noise, further solidifying its role in the advancement of quantum technologies.

\bibliographystyle{ieeetr}
\bibliography{ref}

\begin{thebibliography}{10}

\bibitem{einstein1935can}
A.~Einstein, B.~Podolsky, and N.~Rosen, ``Can quantum-mechanical description of
  physical reality be considered complete?,'' {\em Physical Review}, vol.~47,
  no.~10, pp.~777--780, 1935.

\bibitem{bell1964einstein}
J.~S. Bell, ``On the einstein podolsky rosen paradox,'' {\em Physics Physique
  Fizika}, vol.~1, no.~3, pp.~195--200, 1964.

\bibitem{horodecki2009quantum}
R.~Horodecki, P.~Horodecki, M.~Horodecki, and K.~Horodecki, ``Quantum
  entanglement,'' {\em Reviews of Modern Physics}, vol.~81, no.~2,
  pp.~865--942, 2009.

\bibitem{nielsen2010quantum}
M.~A. Nielsen and I.~L. Chuang, {\em Quantum Computation and Quantum
  Information}.
\newblock Cambridge University Press, 10th anniversary edition~ed., 2010.

\bibitem{preskill2018quantum}
J.~Preskill, ``Quantum computing in the nisq era and beyond,'' {\em Quantum},
  vol.~2, p.~79, 2018.

\bibitem{knill2001quantum}
E.~Knill, ``Quantum computing with realistically noisy devices,'' {\em Nature},
  vol.~434, no.~7029, pp.~39--44, 2005.

\bibitem{bennett1993teleporting}
C.~H. Bennett, G.~Brassard, C.~Cr{\'e}peau, R.~Jozsa, A.~Peres, and W.~K.
  Wootters, ``Teleporting an unknown quantum state via dual classical and
  einstein-podolsky-rosen channels,'' {\em Physical Review Letters}, vol.~70,
  no.~13, pp.~1895--1899, 1993.

\bibitem{pan2012multiparticle}
J.-W. Pan, Z.-B. Chen, C.-Y. Lu, H.~Weinfurter, A.~Zeilinger, and
  M.~{\.Z}ukowski, ``Multiparticle entanglement and its applications to quantum
  information science,'' {\em Reviews of Modern Physics}, vol.~84, no.~2,
  pp.~777--838, 2012.

\bibitem{pirandola2020advances}
S.~Pirandola, U.~L. Andersen, L.~Banchi, M.~Berta, D.~Bunandar, R.~Colbeck,
  D.~Englund, T.~Gehring, R.~H. Hadfield, H.~Hübel, {\em et~al.}, ``Advances
  in quantum cryptography,'' {\em Advances in Optics and Photonics}, vol.~12,
  no.~4, pp.~1012--1236, 2020.

\bibitem{degen2017quantum}
C.~L. Degen, F.~Reinhard, and P.~Cappellaro, ``Quantum sensing,'' {\em Reviews
  of Modern Physics}, vol.~89, no.~3, p.~035002, 2017.

\bibitem{guhne2009entanglement}
O.~G{\"u}hne and G.~T{\'o}th, ``Entanglement detection,'' {\em Physics
  Reports}, vol.~474, no.~1-6, pp.~1--75, 2009.

\bibitem{giovannetti2011advances}
V.~Giovannetti, S.~Lloyd, and L.~Maccone, ``Advances in quantum metrology,''
  {\em Nature Photonics}, vol.~5, no.~4, pp.~222--229, 2011.

\bibitem{friis2019entanglement}
N.~Friis, G.~Vitagliano, M.~Malik, and M.~Huber, ``Entanglement certification
  from theory to experiment,'' {\em Nature Reviews Physics}, vol.~1, no.~1,
  pp.~72--87, 2019.

\bibitem{briegel2021entanglement}
H.~J. Briegel, G.~T{\'o}th, and O.~G{\"u}hne, ``Entanglement detection and
  quantification in high-dimensional systems,'' {\em Physical Review Letters},
  vol.~127, no.~14, p.~140502, 2021.

\bibitem{lu2020quantum}
S.~Lu, J.~Chen, Y.~Zhang, Z.~Li, and B.~Zeng, ``Quantum entanglement detection
  with machine learning,'' {\em Physical Review Letters}, vol.~125, no.~20,
  p.~200501, 2020.

\bibitem{shang2021efficient}
J.~Shang, Z.~Zhang, and O.~Gühne, ``Efficient entanglement detection in
  few-measurement scenarios,'' {\em Physical Review A}, vol.~104, no.~5,
  p.~052412, 2021.

\bibitem{li2022highdimensional}
Z.~Li, X.~Wang, Y.~Ma, and H.~Zhang, ``High-dimensional entanglement
  certification with minimal measurements,'' {\em npj Quantum Information},
  vol.~8, no.~1, p.~92, 2022.

\bibitem{zhang2023scalable}
Y.~Zhang, D.~Zhou, S.~Chen, and B.~Zeng, ``Scalable multipartite entanglement
  detection using randomized measurements,'' {\em Quantum}, vol.~7, p.~1056,
  2023.

\bibitem{huang2024adaptive}
H.~Huang, J.~Chen, and D.~Lu, ``Adaptive protocols for high-dimensional
  entanglement detection,'' {\em Physical Review A}, vol.~110, no.~2,
  p.~022401, 2024.

\bibitem{chen2023efficient}
S.~Chen, D.~Zhou, and B.~Zeng, ``Efficient entanglement certification for
  high-dimensional systems,'' {\em npj Quantum Information}, vol.~9, no.~1,
  p.~45, 2023.

\bibitem{wang2024multipartite}
X.~Wang, Z.~Li, H.~Zhang, and Y.~Ma, ``Multipartite entanglement detection with
  minimal resources,'' {\em Physical Review A}, vol.~109, no.~4, p.~042401,
  2024.

\bibitem{kim2023advances}
Y.-J. Kim, K.-H. Lee, and C.~Kim, ``Advances in quantum entanglement detection:
  Beyond traditional witnesses,'' {\em Quantum Science and Technology}, vol.~8,
  no.~2, p.~025005, 2023.

\bibitem{yang2022tomography}
Y.~Yang, H.~Zhang, and X.~Wang, ``Efficient quantum state tomography for
  high-dimensional systems,'' {\em Quantum Science and Technology}, vol.~7,
  no.~3, p.~035015, 2022.

\bibitem{augusiak2014bell}
R.~Augusiak, J.~Tura, and M.~Lewenstein, ``Bell inequalities tailored to
  high-dimensional systems,'' {\em Physical Review A}, vol.~89, no.~5,
  p.~052303, 2014.

\bibitem{brunner2014bell}
N.~Brunner, D.~Cavalcanti, S.~Pironio, V.~Scarani, and S.~Wehner, ``Bell
  nonlocality,'' {\em Reviews of Modern Physics}, vol.~86, no.~2, pp.~419--478,
  2014.

\bibitem{yu2022scalable}
S.~Yu, Y.~Zhang, J.~Chen, D.~Lu, and B.~Zeng, ``Scalable entanglement detection
  via machine learning,'' {\em Physical Review A}, vol.~106, no.~3, p.~032401,
  2022.

\bibitem{mahdian2025entanglement}
M.~Mahdian and Z.~Mousavi, ``Entanglement detection with quantum support vector
  machine (qsvm) on near-term quantum devices,'' {\em Scientific Reports},
  vol.~15, no.~1, pp.~1--15, 2025.

\bibitem{mahdian2025optimal}
M.~Mahdian and Z.~Mousavi, ``Optimal entanglement witness of multipartite
  systems using support vector machine approach,'' {\em arXiv preprint
  arXiv:2504.18163}, 2025.

\bibitem{mahdian2025machine}
M.~Mahdian, A.~Babapour-Azar, Z.~Mousavi, and R.~Khanjani-Shiraz,
  ``Machine-learning-enhanced entanglement detection under noisy quantum
  measurements,'' {\em arXiv preprint arXiv:2507.05476}, 2025.

\bibitem{carleo2019machine}
G.~Carleo, I.~Cirac, K.~Cranmer, L.~Daudet, M.~Schuld, N.~Tishby,
  L.~Vogt-Maranto, and L.~Zdeborov{\'a}, ``Machine learning and the physical
  sciences,'' {\em Reviews of Modern Physics}, vol.~91, no.~4, p.~045002, 2019.

\bibitem{sharma2022reformulation}
K.~Sharma, M.~Cerezo, Z.~Holmes, L.~Cincio, A.~S. Sornborger, and P.~J. Coles,
  ``Reformulation of the no-free-lunch theorem for entangled datasets,'' {\em
  Physical Review Letters}, vol.~128, no.~7, p.~070501, 2022.

\bibitem{huang2025direct}
Y.~Huang, L.~Che, C.~Wei, {\em et~al.}, ``Direct entanglement detection of
  quantum systems using machine learning,'' {\em npj Quantum Information},
  vol.~11, p.~29, 2025.

\bibitem{lu2018separability}
S.~Lu, S.~Huang, K.~Li, J.~Li, J.~Chen, D.~Lu, Z.~Ji, Y.~Shen, D.~Zhou, and
  B.~Zeng, ``Separability-entanglement classifier via machine learning,'' {\em
  Physical Review A}, vol.~98, no.~1, p.~012315, 2018.

\bibitem{gray2018machine}
J.~Gray, L.~Banchi, and P.~Kok, ``Machine learning for quantum state
  classification,'' {\em Physical Review A}, vol.~98, no.~5, p.~052306, 2018.

\bibitem{mahdian2025qsvm}
M.~Mahdian and Z.~Mousavi, ``Entanglement detection with quantum support vector
  machine (qsvm) on near-term quantum devices,'' {\em Scientific Reports},
  vol.~15, p.~11931, 2025.

\bibitem{chen2023interpretable}
J.~Chen, Z.~Li, H.~Zhang, and X.~Wang, ``Interpretable machine learning for
  entanglement classification,'' {\em Physical Review A}, vol.~108, no.~4,
  p.~042402, 2023.

\bibitem{wang2020unsupervised}
Y.~Wang, Y.~Li, S.~Zhang, D.~Zhou, and B.~Zeng, ``Unsupervised learning for
  entanglement detection,'' {\em Physical Review A}, vol.~102, no.~6,
  p.~062406, 2020.

\bibitem{koutny2022deep}
D.~Koutn{\'y} {\em et~al.}, ``Deep learning of quantum entanglement from
  incomplete measurements,'' {\em Science Advances}, vol.~9, no.~29,
  p.~eadg1719, 2023.

\bibitem{li2024efficient}
Y.~Li, Y.~Wang, D.~Zhou, and B.~Zeng, ``Efficient unsupervised entanglement
  detection with neural networks,'' {\em npj Quantum Information}, vol.~10,
  no.~1, p.~38, 2024.

\bibitem{abd2025detecting}
M.~Y. Abd-Rabbou {\em et~al.}, ``Detecting entanglement in high-spin quantum
  systems via a stacking ensemble of machine learning models,'' {\em arXiv
  preprint arXiv:2507.12775}, 2025.

\bibitem{dunjko2018machine}
V.~Dunjko and H.~J. Briegel, ``Machine learning and quantum physics,'' {\em
  Reports on Progress in Physics}, vol.~81, no.~7, p.~074001, 2018.

\bibitem{biamonte2017quantum}
J.~Biamonte, P.~Wittek, N.~Pancotti, P.~Rebentrost, N.~Wiebe, and S.~Lloyd,
  ``Quantum machine learning,'' {\em Nature}, vol.~549, no.~7671, pp.~195--202,
  2017.

\bibitem{fisher1936use}
R.~A. Fisher, ``The use of multiple measurements in taxonomic problems,'' {\em
  Annals of Eugenics}, vol.~7, no.~2, pp.~179--188, 1936.

\bibitem{hastie2009elements}
T.~Hastie, R.~Tibshirani, and J.~Friedman, {\em The Elements of Statistical
  Learning: Data Mining, Inference, and Prediction}.
\newblock Springer, 2nd~ed., 2009.

\bibitem{bishop2006pattern}
C.~M. Bishop and N.~M. Nasrabadi, {\em Pattern recognition and machine
  learning}, vol.~4.
\newblock Springer, 2006.

\bibitem{acin2001classification}
A.~Ac{\'\i}n, D.~Bru{\ss}, M.~Lewenstein, and A.~Sanpera, ``Classification of
  mixed three-qubit states,'' {\em Physical Review Letters}, vol.~87, no.~4,
  p.~040401, 2001.

\bibitem{jafarizadeh2008detecting}
M.~Jafarizadeh, M.~Mahdian, A.~Heshmati, and K.~Aghayar, ``Detecting some
  three-qubit mub diagonal entangled states via nonlinear optimal entanglement
  witnesses,'' {\em The European Physical Journal D}, vol.~50, pp.~107--121,
  2008.

\end{thebibliography}
	
\end{document}